\documentclass[12pt]{article}

\usepackage{latexsym,axodraw}

\def\al{\alpha^{\prime}}
\def\a{& \hspace{-7pt}}

\newcommand{\eqalign}[1]
{\hspace{-10pt}\begin{array}{ll}#1\end{array}\hspace{-10pt}}

\def\bea{\begin{eqnarray}}
\def\eea{\end{eqnarray}}
\def\be{\begin{equation}}
\def\ee{\end{equation}}
\def\nn{\nonumber}
\def\a{& \hspace{-7pt}}
\def\c{\hspace{-5pt}}
\def\Z{{\bf Z}}
\def\5{\bar 5}

\begin{document}

\thispagestyle{empty}

\begin{center}
\hfill CERN-TH/2002-063\\
\hfill SISSA-20/2002/EP \\

\begin{center}

\vspace{1.7cm}

{\Large\bf Open string models with Scherk-Schwarz \\[2mm]
SUSY breaking and localized anomalies}

\end{center}

\vspace{1.4cm}

{\sc C.A. Scrucca$^{a}$, M. Serone$^{b}$ and M. Trapletti$^{b}$}\\

\vspace{1.2cm}

${}^a$
{\em CERN, 1211 Geneva 23, Switzerland}
\vspace{.3cm}

${}^b$
{\em ISAS-SISSA, Via Beirut 2-4, 34013 Trieste, and INFN, Trieste, Italy}
\vspace{.3cm}

\end{center}

\vspace{0.8cm}

\centerline{\bf Abstract}
\vspace{2 mm}
\begin{quote}\small
We study examples of chiral four-dimensional IIB orientifolds with
Scherk--Schwarz supersymmetry breaking, based on freely acting
orbifolds. We construct a new $\Z_3\times \Z_3^\prime$ model,
containing only $D9$-branes, and rederive from a more geometric
perspective the known $\Z_6^\prime\times \Z_2^\prime$ model,
containing $D9$, $D5$ and $\bar D 5$ branes. The cancellation of
anomalies in these models is then studied locally in the internal
space. These are found to cancel through an interesting generalization
of the Green--Schwarz mechanism involving twisted Ramond--Ramond axions
and 4-forms. The effect of the latter amounts to local counterterms
from a low-energy effective field theory point of view.
We also point out that the number of spontaneously broken $U(1)$ gauge
fields is in general greater than what expected from a four-dimensional
analysis of anomalies.
\end{quote}

\vfill

\newpage
\setcounter{equation}{0}

\section{Introduction}

Supersymmetry (SUSY) is certainly one of the key ideas to understand how to embed the
Standard Model (SM) into a more fundamental microscopic theory. It provides, among other
things, an elegant solution to the hierarchy problem (stabilizing the electroweak scale)
if broken at a sufficiently low scale $M_{\rm susy}\sim {\rm TeV}$. In string theory, SUSY
plays an even more important role and apparently represents a crucial ingredient in defining
absolutely consistent models. In fact, the construction of truly stable non-SUSY string
vacua is tremendously hard and no such model has been found so far. While waiting for new principles
or breakthroughs that hopefully will shed light on this fundamental problem, it is
nevertheless very important to explore the structure and main properties of SUSY breaking
in string theory.

One of the most interesting and promising mechanisms of symmetry breaking in theories with
compact extra-dimensions, such as string theory, is the so-called Scherk--Schwarz (SS) symmetry-breaking 
mechanism \cite{SS}, which consists in suitably twisting the periodicity conditions
of each field along some compact directions. In this way, one obtains a non-local, perturbative
and calculable symmetry-breaking mechanism. String models of this type can be constructed by
deforming supersymmetric orbifold \cite{orb} models, and a variety of four-dimensional (4D)
closed string models, mainly based on $\Z_2$ orbifolds, have been constructed in this way
\cite{ssstringa}. More in general, SS symmetry breaking can be achieved through freely acting
orbifold projections \cite{kk}. This fact has recently been exploited in \cite{root} to construct
a novel class of closed string examples, including a model based on the $\Z_3$ orbifold.
Unfortunately, a low compactification scale is quite unnatural for closed string models, where
the fundamental string scale $M_s$ is tied to the Planck scale, and can be achieved only in very
specific situations \cite{large} (see also \cite{amq}). The situation is different for open strings,
where $M_s$ can be very low \cite{mill}, and interesting open string models with SS SUSY breaking
have been derived in \cite{ads1,adds,cotrone}. Recently, the SS mechanism has been object of renewed
interest also from a more phenomenological ``bottom-up'' viewpoint, where it has been used in
combination with orbifold projections to construct realistic 5D non-SUSY extensions of the SM
\cite{pom,bhn}.

The main aim of this paper is to exploit the general ideas proposed in \cite{root} to construct
chiral IIB compact orientifold models with SS supersymmetry breaking.
We derive a new $\Z_3 \times \Z_3^\prime$ orientifold by applying a freely acting $\Z_3^\prime$
projection defined as a translation of order 3 and a non-SUSY twist to the known SUSY
$\Z_3$ orientifold \cite{abpss,afiv}. The model turns out to be chiral and extremely simple, since
only $D9$-branes are present. It exhibits SS SUSY breaking in both the closed and open string sectors.
All the gauginos are massive, but there is an anomalous spectrum of massless charginos.
The model is classically stable, since all massless Neveu--Schwarz--Neveu--Schwarz (NSNS)
and Ramond--Ramond (RR) tadpoles vanish, and potential tachyons can be avoided by taking
a sufficiently large volume for the SS torus, i.e. the torus where the translation acts.
We also rederive from a more geometrical perspective the $\Z_6^\prime \times \Z_2^\prime$ model of
\cite{adds} (see also \cite{antoniadis}), by applying to the SUSY $\Z_6^\prime$ model of
\cite{afiv} a freely acting $\Z_2^\prime$ projection generated by a translation of order 2 along
a circle combined with a $(-)^F$ operation, where $F$ is the 4D space-time fermion number operator.
We then discuss in some detail its rich structure involving $D9$, $D5$ and $\bar D5$ branes.

An other important goal of this work is to perform a detailed study of {\it local}
anomaly cancellation (i.e. point-by-point in the compact space) for this kind of models.
This study is motivated by the results of \cite{sssz} where it has been pointed out
that orbifold field theories can have anomalies localized at fixed points that vanish
when integrated over the internal space, and originate from loops of heavy Kaluza--Klein (KK) modes.
To this aim, we will extend the approach that has been followed in \cite{ssd4} for 4D SUSY orientifolds
to distinguish between different points in the internal space. We find that all anomalies cancel locally,
thanks to an interesting Green--Schwarz (GS) mechanism \cite{GS} involving twisted RR axions belonging
to 4D sectors localized at fixed points or 6D sectors localized at fixed-planes, 
as found in \cite{iru,ssd4}, but also 4-forms coming from 6D
sectors localized at fixed planes. The latter effect arises whenever RR tadpoles are cancelled
globally but not locally\footnote{The global cancellation of RR tadpoles ensures only the global
cancellation of cubic irreducible anomalies; see {\rm e.g.} \cite{tad/ano}.},
and involves only heavy KK modes of the 4-forms.
In non-compact string vacua, such as intersecting branes, this kind of effect is already
included in the usual anomaly inflow of \cite{ghm}.
Global irreducible anomalies can arise in this case, since there is no constraint on the global RR flux;
they are cancelled thanks to RR forms propagating in more than 4D.
This shows once again the very close relation between
the GS mechanism and the inflow mechanism of \cite{ch}, even for irreducible terms.

Our results reveal an important distinction between anomalies appearing through a 6-form in the
anomaly polynomial and anomalies appearing through the product of a 2-form and a 4-form.
In the former case, the GS mechanism is mediated by twisted RR 4-forms and the
corresponding symmetry is linearly realized. In the latter, instead, anomalies are cancelled
by a GS mechanism mediated by twisted RR axions, and the symmetry is realized only non-linearly.
When applied to a $U(1)$ factor with an anomaly that is globally but not locally vanishing, these
two situations lead respectively to a massless and massive 4D photon\footnote{As in the standard
case \cite{dsw}, a pseudo-anomalous photon can become massive by eating an axion through a Higgs
mechanism.}. This leads to the important conclusion that the number of spontaneously broken $U(1)$
gauge factors is in general {\it greater} than what is expected from a global analysis of anomalies.
This fact, which has not been appreciated so far in the literature, could have an important impact
in the context of open string phenomenology. The difference between the two mechanisms
involving axions and 4-forms is particularly striking from a 4D low-energy effective field theory point of
view, where heavy KK modes are integrated out. The axions remain dynamical, but the 4-forms must
be integrated out, and we will show that their net effect then amounts to a local 6D Chern--Simons
counterterm with a discontinuous coefficient, jumping at the fixed points; this counterterm 
thus occurs in a way that is manifestly compatible with local supersymmetry
and falls in the category of terms discussed in \cite{bkp} (see also \cite{hw}).
This realizes a 6D version of the possibility of cancelling globally vanishing anomalies
through a dynamically generated Chern--Simons term \cite{sssz}.
 It also confirms in a string context that operators that are odd under
the orbifold projection can and do in general occur in the 4D effective theory with odd
coefficients, as emphasized in \cite{bccrs}.

The paper is organized as follows. In section 2 we rederive the model presented in
\cite{adds} as a $\Z_6^\prime \times \Z_2^\prime$ orientifold, emphasizing geometrical
aspects. In section 3, the novel $\Z_3\times \Z_3^\prime$ model is constructed and
described in some detail.  Section 4 is devoted to the study of local anomaly cancellation,
and contains both a general discussion and a detailed analysis for the two models at hand.
In section 5 we state our conclusions. Finally, some more details concerning lattice
sums and anomalous couplings are reported in two appendices, where we also clarify an issue
left partially unsolved in \cite{ssd4}, regarding the factorization of anomalous
couplings in twisted sectors with fixed planes.

\section{\label{Z6} The $\Z_6^\prime \times \Z_2^\prime$ model}

The $\Z_6^\prime \times \Z_2^\prime$ orientifold of \cite{adds} is
obtained by applying a SUSY-breaking $\Z_2^\prime$ projection to the
SUSY $\Z_6^\prime$ model of \cite{afiv}. The $\Z_6^\prime$ group is generated
by $\theta$, acting as rotations of angles $2\pi v^\theta_i$ in the  three
internal tori $T^2_i$ ($i=1,2,3$), with $v^\theta_i=1/6 (1,-3,2)$. The
$\Z_2^\prime$ group is instead generated by $\beta$, acting as a translation
of length $\pi R$ along one of the radii of $T^2_2$ (that we shall call SS
direction in the following), combined with a sign $(-)^F$, where $F$ is the
4D space-time fermion number. Beside the $O9$-plane, the model contains $O5$-planes
at $y=0$ and $y=\pi R$ along the SS direction (as the corresponding
SUSY model \cite{afiv}) and $\bar O5$-planes
at $y=\pi R/2$ and $y=3\pi R/2$ along the SS direction (see Figs.~1 and 2),
corresponding to the two elements of order 2, $\theta^3$ and $\theta^3\beta$.
In order to cancel both NSNS and RR massless tadpoles, $D9$, $D5$ and $\bar D 5$-branes
must be introduced.

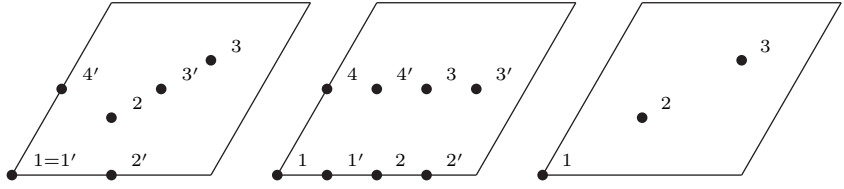
\begin{figure}[h]
\begin{center}
\begin{picture}(280,85)(0,0)
\Line(0,0)(75,0)
\Line(37.5,65)(112.5,65)
\Line(0,0)(37.5,65)
\Line(75,0)(112.5,65)
\Vertex(0,0){2}
\put(8,5){\footnotesize ${{}_{1=1^\prime}}$}
\Vertex(37.5,21.65){2}
\put(45.5,26.65){\footnotesize${ {}_2}$}
\Vertex(75,43.3){2}
\put(83, 48.3){\footnotesize${ {}_3}$}
\Vertex(37.5,0){2}
\put(45,5){\footnotesize${ {}_{2^\prime}}$}
\Vertex(56.25,32.48){2}
\put(64.25,37.48){\footnotesize${ {}_{3^\prime}}$}
\Vertex(18.75,32.48){2}
\put(26.75,37.48){\footnotesize${ {}_{4^\prime}}$}
\Line(200,0)(275,0)
\Line(237.5,65)(312.5,65)
\Line(200,0)(237.5,65)
\Line(275,0)(312.5,65)
\Vertex(200,0){2}
\put(208,5){\footnotesize${ {}_1}$}
\Vertex(237.5,21.65){2}
\put(245.5,26.65){\footnotesize${ {}_2}$}
\Vertex(275,43.3){2}
\put(283, 48.3){\footnotesize${ {}_3}$}
\Line(100,0)(175,0)
\Line(137.5,65)(212.5,65)
\Line(100,0)(137.5,65)
\Line(175,0)(212.5,65)
\Vertex(100,0){2}
\put(108,5){\footnotesize${ {}_1}$}
\Vertex(137.5,0){2}
\put(145,5){\footnotesize${ {}_2}$}
\Vertex(156.25,32.48){2}
\put(164.25,37.48){\footnotesize${ {}_3}$}
\Vertex(118.75,32.48){2}
\put(126.75,37.48){\footnotesize${ {}_4}$}
\Vertex(118.75,0){2}
\put(126.75,5){\footnotesize${ {}_{1^\prime}}$}
\Vertex(156.25,0){2}
\put(164.25,5){\footnotesize${ {}_{2^\prime}}$}
\Vertex(175,32.48){2}
\put(183,37.48){\footnotesize${ {}_{3^\prime}}$}
\Vertex(137.5,32.48){2}
\put(145.5,37.48){\footnotesize${ {}_{4^\prime}}$}
\end{picture}
\caption{\label{fix1}
{\footnotesize \em  The fixed-points structure in the $\Z_6^\prime \times \Z_2^\prime$ model.
We label the 12 $\theta$ fixed points with $P_{1bc}$ and the 12 $\theta\beta$ fixed
points with $P_{1bc^\prime}$, each index referring to a $T^2$, ordered as in the figure.
Similarly, we denote with
$P_{a\bullet c}$ the 9 $\theta^2$ fixed planes filling the second $T^2$, and
respectively with
$P_{a^\prime b\bullet }$ and $P_{a^\prime  b^\prime \bullet}$ the 16 $\theta^3$ fixed
and $\theta^3\beta$ fixed planes filling the third $T^2$. The 32 $D\mathit{5}$-branes
and the 32 $\bar D \mathit{5}$-branes are located at point 1 in the first $T^2$, fill the
third $T^2$, and sit at the points $1$ and $1^\prime$ respectively in the second $T^2$.}}
\vskip -10pt
\end{center}
\end{figure}

\subsection{Closed string spectrum}

The main features of the closed string spectrum of the $\Z_6^\prime \times \Z_2^\prime$
model can be deduced from those of the $\Z_6^\prime$ model, which can be found in \cite{afiv}.
The only SUSY-breaking generators are $\beta$, $\theta^2\beta$ and $\theta^4\beta$;
all the other elements preserve some SUSY (generically different from sector to sector).
The $\Z_2^\prime$ projection acts therefore in a SUSY-breaking way in the untwisted and
$\theta^{2,4}$ twisted sectors, and in a SUSY-preserving way in the
remaining $\theta^{1,5}$ and $\theta^3$ twisted sectors of the $\Z_6^\prime$ model.
In addition, we must consider the new $\theta^k\beta$ twisted sectors.

Consider first the $\theta^k$ sectors already present in the $\Z_6^\prime$ model.
In the untwisted sector,
one gets a gravitational multiplet and 5 chiral multiplets of $N\!=\! 1$ SUSY, and the $\Z_2^\prime$
projection eliminates all the fermions. In the $\theta^{2,4}$ twisted sectors, one
gets 9 hypermultiplets of $N\!=\! 2$ SUSY, and the $\Z_2^\prime$ projection again eliminates
all the fermions.
Finally, the $\theta^{1,5}$ and $\theta^3$ twisted sectors give each 12 chiral multiplets of
$N\!=\! 1$ SUSY, and the $\Z_2^\prime$ action reduces this number to 6, since it identifies sectors
at fixed points that differ by a $\pi R$ shift in the position along the SS direction.

Consider next the new $\theta^k\beta$ sectors emerging in the $\Z_6^\prime \times \Z_2^\prime $
model. The $\beta$ twisted sector yields one real would-be tachyon of mass $\alpha'm^2=-2+ R^2/(2\al)$.
Similarly, the $\theta^2 \beta$ sectors yield 6 complex would-be tachyons of mass
$\alpha'm^2=-2/3+ R^2/(2\al)$\footnote{These are clearly the lightest would-be tachyons
in both the $\beta$ and $\theta^2\beta$ twisted sectors, but it should be recalled that there
is actually an infinite tower of such states, with increasing winding mode.}.
Finally, the $\theta^{1,5}\beta$ twisted and $\theta^{3}\beta$ twisted sectors each give
6 chiral multiplets of $N\!=\! 1^\prime$ SUSY, which have opposite chirality and a different unbroken
SUSY compared to those arising in the $\theta^{1,5}$ and $\theta^{3}$-sectors, because $\beta$
changes the GSO projection due to the $(-1)^F$ operation that it involves.

The closed string spectrum that we have just derived is summarized for convenience in
Table \ref{closedspectrum}.

\subsection{Tadpole cancellation}

As mentioned above, the $\Omega$-projection in the closed string sector
introduces $O9$, $O5$, $\bar O 5$ planes and hence a non-vanishing number
$n_9$, $n_5$ and $n_{\bar 5}$ of $D9$, $D5$ and $\bar D 5$ branes is needed
to cancel all massless tadpoles.

The computation of the partition functions on the annulus ($A$), M\"obius strip ($M$)
and Klein bottle ($K$) surfaces and the extraction of the tadpoles is standard, although
lengthy, and we do not report all the details. The world-sheet parity operator $\Omega$
is defined in such a way that $\Omega \Phi (\sigma) \Omega^{-1} = \Phi (2\pi - \sigma)$
for a generic world-sheet field $\Phi(\sigma)$, and its action on the RR
and NSNS vacua is given by $\Omega|0\rangle_{NSNS}=-|0\rangle_{NSNS}$ and
$\Omega|0\rangle_{RR}=-|0\rangle_{RR}$ in the closed string sector and
by $\Omega|0\rangle_{NS}=-i |0\rangle_{NS}$ and $\Omega|0\rangle_{R}=-|0\rangle_{R}$
in the open string sector. The modular parameter for the $A$, $M$ and $K$ surfaces is
taken to be $t_A = it$, $t_M = it-1/2$ and $t_K = 2it$.
The modular transformation needed to switch from the direct to the transverse channel
is simply $S:\tau\rightarrow -1/\tau$ for the $A$ and $K$ surfaces. For the $M$
surface, the appropriate transformation is instead $P=TST^2ST$, where
$T:\tau\rightarrow\tau+1$. The corresponding modular parameters in the
transverse channel are $l_A=1/(2t)$, $l_M=1/(8t)$ and $l_K=1/(4t)$.

The only novelty with respect to the $\Z_6^\prime$ model are the non-SUSY sectors
that arise when the $\Z_2^\prime$ generator $\beta$ enters as twist or insertion in
the trace defining the partition function. The corresponding contributions to the
partition functions can be easily deduced from their analogues in the $\Z_6^\prime$
model. In the $K$ amplitude, owing to the presence of $\Omega$, the insertion
of $\beta$ acts only in the lattice contribution, as reported in eq.~(\ref{Klatt}).
As a twist, $\beta$ inverts the GSO projection and acts in the lattice.
This implies that the $\beta$ twisted contribution,
after the $S$ modular transformation, will be the same as the SUSY
untwisted sector contribution, but proportional to $(1_{NSNS}+1_{RR})$
instead of the usual $(1_{NSNS}-1_{RR})$, and with some terms dropped due
to the vanishing of the lattice contribution as in (\ref{Klatt}).
This represents a non-vanishing tadpole for the untwisted RR six-form,
and reflects the presence of $\bar O 5$-planes (beside $O5$-planes) in this model.
On the $A$ and $M$ surfaces, the insertion of $\beta$ acts in the lattices
as discussed in Appendix A. Apart from that, it simply reverts the R contribution
to the partition function. This simple sign flip has, however, different consequences
in the two surfaces when analysing the closed string channel, because of the two different
modular transformations ($S$ and $P$) that are involved. For the $M$ amplitude, the result
is obtained from its SUSY analogue by replacing the factor $(1_{NSNS}-1_{RR})$ by
$(1_{NSNS}+1_{RR})$, and has a clear interpretation as $D$-branes/$\bar O$-planes and
$\bar D$-branes/$O$-planes interactions. For the $A$ amplitude, the action of $\beta$ in the
closed string channel reverses the GSO projection and, depending on which $\Z_6^\prime$ generator
is inserted (and which boundary conditions are considered), this can lead to an exchange of 
would-be tachyons.

The group action on the Chan--Paton degrees of freedom is encoded in the twist
matrices $\gamma$ and $\delta$, respectively for the $\Z_6^\prime$ and
$\Z_2^\prime$ generators.
The group algebra, as usual, allows us to write the Chan--Paton contribution of a $M$ amplitude
with the insertion of $\theta^n\beta^m$ as $ \pm {\bf Tr} (\gamma^n\delta^m)^2$,
the freedom of sign being fixed by tadpole cancellation and by the
relative action of $\Omega$ on 5 and 9 branes, as studied by Gimon
and Polchinski (see for example \cite{GP,afiv}).
Tadpole cancellation and the $\Omega$ action fix $\gamma^6 = -I$ in
the 9, 5 and $\5$ sector, as in \cite{afiv}, and $\delta^2 = -I$ in the 5 and
$\bar 5$ sectors, with the further condition $\{\gamma,\delta\}=0$.
We also impose $\delta^2=I$ in the 9 sector; the case  $\delta^2=-I$ will be
considered later on.

To be fully general, we will use, for the twist matrices $\gamma$
in the $5$ and $\bar 5$ sectors, an extra index that distinguishes between distinct $\theta^k$ fixed
points (or fixed planes). Similarly, an other extra index is needed also for the
matrices $\gamma\delta$ in the $5$ and $\bar 5$ sectors, running over the
$\theta^k\beta$ fixed points.

The final form of the massless tadpoles is most conveniently presented by
distinguishing the two closed string sectors with a sign $\eta$ equal to
$+1$ for the NSNS sector and $-1$ for the RR sector. The result is given
by $v_4/12\int \!dl$ times
\bea
I:\a\a
\frac{v_1v_2v_3}{8}\,
\eta \left[2^5-n_9\right]^2 + \frac{v_3}{8v_1v_2}\,
\eta \left[2^6 \delta_{\eta,1}-n_5 - \eta\,n_{\5}\right]^2
\label{firsttad} \;, \\
\theta:\a\a
\frac{\sqrt{3}}{6}\sum_{c=1}^3
\sum_{b=1}^4 \eta \left[2^{-1}\,{\bf Tr}\,\gamma_{9}
-{\bf Tr}\,\gamma_{5b} - \eta\,{\bf Tr}\,\gamma_{\5b}\right]^2 \;, \label{tad2Z6} \\
\theta\beta:\a\a
\frac{\sqrt{3}}{6}\sum_{c=1}^3
\sum_{b^\prime=1}^4 \eta \left[2^{-1}\,{\bf Tr}\,\gamma_{9}\delta_{9}
-\eta\,{\bf Tr}\,\gamma_{5b^\prime}\delta_{5}
-{\bf Tr}\,\gamma_{\5b^\prime}\delta_{\5}\right]^2 \;, \label{tad3Z6} \\
\theta^2:\a\a
\frac{1}{4v_2}\sum_{a,c=1}^3
\eta \left[2^4\,\delta_{a,1}\,\delta_{\eta,1} + {\bf Tr}\,\gamma^2_{5ac}
+ \eta\,{\bf Tr}\,\gamma^2_{\5ac}\right]^2
+ \frac{v_2}{12}\sum_{a,b=1}^{3}
\eta \left[2^3+{\bf Tr}\,\gamma^2_{9}\right]^2 \;, \label{tad4Z6} \\
\theta^3:\a\a
v_3\,
\sum_{a^\prime,b=1}^{4} \eta \left[2^{-2}\,{\bf Tr}\,\gamma^3_{9}
+{\bf Tr}\,\gamma^3_{5b}+\eta\,{\bf Tr}\,\gamma^3_{\5b}\right]^2 \;,\label{tad5Z6} \\
\theta^3\beta:\a\a
v_3\,
\sum_{a^\prime,b^\prime=1}^{4} \eta \left[\,2^{-2}\,{\bf Tr}\,\gamma^3_{9}\delta_{9}
+\eta\,{\bf Tr}\,\gamma^3_{5b^\prime}\delta_{5}+{\bf Tr}\,\gamma^3_{\5b^\prime}\delta_{\5}\right]^2
\label{lasttad} \;,
\eea
where we denoted by $\theta^n\beta^m$ the tadpole contribution of the
$\theta^n\beta^m$ twisted closed string states, summed over the various fixed points or planes;
for convenience we have taken the sums
in eqs.~(\ref{tad2Z6}), (\ref{tad3Z6}), (\ref{tad5Z6}) and (\ref{lasttad}) to run
over closed string twisted states and their images under some orbifold elements.
Moreover, $v_4=V_4/(4\pi^2\alpha^\prime)^2$, $v_i=V_i/(4\pi^2\alpha^\prime)$
($i=1,2,3$), with $V_4$ being the volume of the four-dimensional space-time
and $V_i$ the volume of the $T^2_i$.

The NSNS and RR tadpoles differ mainly through relative signs between the contribution
from $D5$ and $\bar D 5$ branes.
In addition, there are cross-cap contributions to the NSNS tadpoles in the $I$ and $\theta^2$
sectors that have no analogue in the RR sector (the terms involving $\delta_{\eta,1}$).

We also report the lightest massive NSNS tadpoles, where would-be tachyons
can develop:
\bea
\beta:\a\a
\frac{v_1v_2v_3}{64}\,q^{-\frac{1}{2}}
\hat\Lambda\left(\frac{1}{2}\right)
\left[{\bf Tr}\,\delta_{9}\right]^2 \;,\label{tadp1} \\
\theta^2\beta:\a\a
\frac{v_2}{24}\,q^{-\frac{1}{6}}
\hat\Lambda\left(\frac{1}{2}\right)
\sum_{a,b=1}^3
\left[{\bf Tr}\,\gamma_{9}^2\delta_{9}\right]^2 \;.
\label{tadp2}
\eea

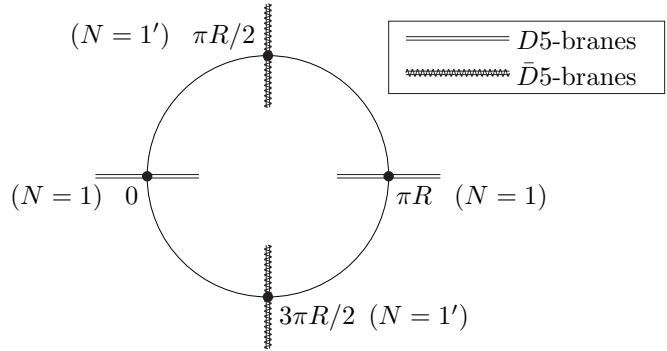
\begin{figure}[h]
\begin{center}
\begin{picture}(330,200)(-80,0)
\setlength{\unitlength}{.65pt}
\SetScale{.65}
\GCirc(100,100){70}{1}
\Vertex(100,30){3}
\Text(107,18)[l]{{\footnotesize $3\pi R/2$\, ($N=1^\prime$)}}
\Vertex(100,170){3}
\Text(93,182)[r]{ \footnotesize ($N=1^\prime$) \, $\pi R/2$}
\Vertex(170,100){3}
\Text(175,88)[l]{\footnotesize$\pi R$ \, ($N=1$)}
\Vertex(30,100){3}
\Text(25,88)[r]{\footnotesize ($N=1$) \, $0$}
\Line(0,99)(60,99)
\Line(0,101)(60,101)
\Line(140,99)(200,99)
\Line(140,101)(200,101)
\Line(101,0)(101,60)
\Line(99,0)(99,60)
\Photon(100,0)(100,60){2}{20}
\Line(101,140)(101,200)
\Line(99,140)(99,200)
\Photon(100,140)(100,200){2}{20}
\EBox(170,190)(330,148)
\Line(180,181)(240,181)
\Line(180,179)(240,179)
\Text(245,180)[l]{\footnotesize$D$5-branes}
\Line(180,161)(240,161)
\Line(180,159)(240,159)
\Photon(180,160)(240,160){2}{20}
\Text(245,160)[l]{\footnotesize$\bar D$5-branes}
\end{picture}
\caption{\label{SSdirection}{\footnotesize \em  Brane positions along the SS
direction for the $\Z_6^\prime\times \Z_2^\prime$ model. The
different supersymmetries left unbroken at the massless level in the $\mathit{55}$ and
$\mathit{\5\!\5}$ sectors are also indicated.}}
\end{center}
\end{figure}

\subsection{Open string spectrum}

We now turn to the determination of a solution for the Chan--Paton matrices satisfying
the above conditions for the global cancellation of massless tadpoles, eqs.~(\ref{firsttad})--(\ref{lasttad}).
For simplicity, we consider the case of maximal unbroken gauge symmetry where
all $D5$ and $\bar D 5$ branes are located respectively at $P_{11\bullet}$ and $P_{11^\prime\bullet}$
(see Fig.~1 and its caption). The $\Z_2^\prime$ projection then requires that an equal number of
image branes be located respectively at $P_{12\bullet}$ and $P_{12^\prime\bullet}$.
We do not consider the case in which branes and antibranes coincide also along the SS direction,
since this configuration is unstable even classically, because of the presence of open string tachyons.
On the other hand, fixing the branes at antipodal points along the SS direction allows a metastable
configuration without open string tachyons for sufficiently large SS radius.

The untwisted tadpoles imply $n_9 = n_5 = n_{\bar 5} = 32$, whereas a definite
solution of the twisted tadpoles is given by
\bea
\gamma_{9} \a=\a \gamma_{5}= \gamma_{\bar 5} \delta_{\bar 5} =
\left(\matrix{\gamma_{16} \a 0 \cr 0 \a - \gamma_{16}}\right) \;;
\label{structure} \\
\delta_{9} \a=\a \left(\matrix{I_{16} \a 0 \cr 0 \a I_{16}}\right) \;,\;\;
\delta_{5} = \delta_{\bar 5} =
\left(\matrix{0 \a I_{16} \cr - I_{16} \a 0}\right)
\label{delta} \;,
\eea
where ($\phi = \exp(i \pi/6)$):
\be
\gamma_{16} = {\rm diag} \{\phi I_2,\phi^5 I_2,\phi^3 I_4,
\overline\phi I_2,\overline\phi^5 I_2,\overline\phi^3 I_4\} \;.
\label{tadsolD4}
\ee
It is easy to verify that with such a choice all massless tadpoles cancel (although
(\ref{tadp1}) and (\ref{tadp2}) do not vanish).
Notice that the above choice for $\gamma_{9,5}$ coincides with that of \cite{afiv}.
The structure of the twist matrices given in (\ref{structure}) and (\ref{delta})
reflects our choice for brane positions; in particular, the matrix $\delta$
implements the translation $\beta$ in the Chan--Paton degrees of freedom. Hence, as
far as the massless spectrum is concerned, we can effectively restrict our attention
to the 16 branes and antibranes at $P_{111}$ and $P_{11^\prime 1}$ respectively, and
work with $16\times 16$ Chan--Paton matrices.

The massless open string spectrum can now be easily derived, and is summarized in Table \ref{spectrum}.
In the 99 sector, the bosonic spectrum is unaffected by the ${\Z_2^\prime}$ element and
therefore coincides with that of the $\Z_6^\prime$ orbifold\footnote{These are as in
\cite{ssd4}, but differ slightly from \cite{afiv} and \cite{antoniadis}.};
all fermions (both gauginos and charginos) are instead massive.
The $55$ and $\5\5$ sectors are supersymmetric at the massless level, but with respect to
different supersymmetries: $N\!=\! 1$ and $N\!=\! 1^\prime$. The $55$ and $\5\5$ gauge groups
$G_5$ and $G_{\bar 5}$ are reduced by the non-trivial action of the translation in
these sectors, and the corresponding states are in conjugate representations. A similar
reasoning also applies for the $95$ and $9\5$ sectors. Finally, the $5\5$ sector does not
contain massless states, thanks to the separations between $D5$-branes and $\bar D 5$-branes.
There are massive scalars and fermions in the bifundamental of $G_5 \times G_{\bar 5}$,
and charged would-be tachyons of mass $\alpha'm^2= -1/2+ R^2/(16\al)$.

Notice that the above solution of the tadpole cancellation conditions is not unique.
In fact, another interesting and more symmetric solution is obtained by choosing
$\delta_9$ of the same form as $\delta_5 = \delta_{\5}$ in (\ref{delta}).
This solution is not maximal in the sense that
the resulting $G_9$ gauge group is reduced and equal to $G_9 = G_5\times G_{\5} = U(4)^2\times U(2)^4$.
However, it has the nice feature that now also the tadpoles (\ref{tadp1}) and (\ref{tadp2})
do vanish. Clearly, there exist other solutions, which we do not report here.
Note for instance that a non-vanishing twist matrix $\delta$ in the 9 sector can be
considered as a $\Z_2$ Wilson line along the SS radius. Since $\delta$
implements a SS gauge symmetry breaking, this reflects the close interplay between
Wilson line symmetry breaking \cite{Wilson-line} and SS gauge symmetry breaking.

Let us now comment on the brane content of this orbifold. {}From the tadpoles, we learn
that there is no local ${\Z_6}$ and ${\Z_2}$ twisted RR charge at all in the model
(${\bf Tr} \gamma = {\bf Tr} \gamma^3 = 0$), but there is a ${\Z_3}$-charge, since
${\bf Tr} \gamma^2 \neq 0$, that globally cancels between $D9$ and $O9$-planes, and $D5$,
$\bar D 5$, $O5$ and $\bar O 5$-planes. On a $\Z_6$ orbifold, a regular $D$-brane must have 5 images.
Since we start with 32 branes, it is clear that the branes in this model cannot all be regular.
In fact, the presence of a non-vanishing $\Z_3$ RR (and NSNS) charge suggests that
fractional $D5$ and $\bar D 5$ branes are present at $\Z_3$ fixed planes\footnote{In
our case, the $\Z_3$ fixed plane is at the origin. However, seen as $D7$-branes wrapped on vanishing
two-cycles \cite{Douglas}, these branes wrap only the $\Z_3$ vanishing cycles.}
of the orbifold. The configuration is then the following. We have 2 regular $D5$ and $\bar D5$
branes (and 5 images for each) and 2 fractional $\Z_3$ $D5$ and $\bar D 5$ branes (and one $\Z_2$
image for each). In our maximal configuration, they are all located at $P_{11\bullet}$ ($D5$) and
$P_{11^\prime\bullet}$ ($\bar D 5$). Clearly, there are the additional $\Z_2^\prime$ images
located at $P_{12\bullet}$ and $P_{12^\prime\bullet}$.
Regular branes can move around freely, whereas fractional branes are stuck at the fixed points.
However, one can still shift a fractional brane from one fixed point to another, suggesting
that this freedom
represents the $T$-dual of discrete Wilson lines in orbifolds. Notice that also $D9$-branes have
${\Z_3}$ RR charge. Although it is not appropriate to speak about fractional $D9$-branes, this
kind of object represents the $T$-dual version of the usual lower-dimensional fractional branes.
In some sense, they are stuck in the gauge bundle, and do not admit continuous Wilson lines, but
only discrete ones.

The $\Z_2^\prime$ twist acts trivially in the gauge-bundle of the $D9$-branes, whereas
in the 5 sector it is $T$-dual to a discrete Wilson line given by the matrix $\delta$.
More precisely, the breaking of the gauge group in the $5$ and $\5$ sector is the $T$-dual
version of a Wilson line symmetry breaking. The additional $(-)^F$ action is on the other hand
responsible for the $D5$ $\rightarrow$ $\bar D 5$ flip for half of the branes.

\section{A $\Z_3 \times \Z_3^\prime$ model}

It has been shown in \cite{root} that SS symmetry breaking can be obtained
also in $\Z_3$ models through a suitable freely acting and SUSY-breaking
$\Z_3^\prime$ projection. In this section, we will construct a new
$\Z_3 \times \Z_3^\prime$ model, based on this structure, that will prove
to be much simpler than the $\Z_6^\prime \times \Z_2^\prime$ model.

The $\Z_3\times\Z^\prime_3$ orbifold group is defined in the following way \cite{root}.
The $\Z_3$ generator $\alpha$ acts as a SUSY-preserving rotation with twist $v^\alpha_i=1/3(1,1,0)$,
while the $\Z_3^\prime$ generator $\beta$ acts as a SUSY-breaking rotation with
$v^\beta_i=1/3(0,0,2)$ and an order-three diagonal translation $\delta$ in $T^2_1$.

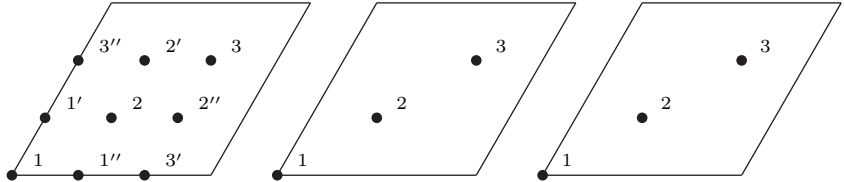
\begin{figure}[h]
\begin{center}
\begin{picture}(280,85)(0,0)
\Line(200,0)(275,0)
\Line(237.5,65)(312.5,65)
\Line(200,0)(237.5,65)
\Line(275,0)(312.5,65)
\Vertex(200,0){2}
\put(208,5){\footnotesize${ {}_1}$}
\Vertex(237.5,21.65){2}
\put(245.5,26.65){\footnotesize${ {}_2}$}
\Vertex(275,43.3){2}
\put(283, 48.3){\footnotesize${ {}_3}$}
\Line(100,0)(175,0)
\Line(137.5,65)(212.5,65)
\Line(100,0)(137.5,65)
\Line(175,0)(212.5,65)
\Vertex(100,0){2}
\put(108,5){\footnotesize${ {}_1}$}
\Vertex(137.5,21.65){2}
\put(145.5,26.65){\footnotesize${ {}_2}$}
\Vertex(175,43.3){2}
\put(183,48.3){\footnotesize${ {}_3}$}
\Line(0,0)(75,0)
\Line(37.5,65)(112.5,65)
\Line(0,0)(37.5,65)
\Line(75,0)(112.5,65)
\Vertex(0,0){2}
\put(8,5){\footnotesize${ {}_1}$}
\Vertex(37.5,21.65){2}
\put(45.5,26.65){\footnotesize${ {}_2}$}
\Vertex(75,43.3){2}
\put(83,48.3){\footnotesize${ {}_3}$}
\Vertex(12.5,21.65){2}
\put(20.5,26.65){\footnotesize${ {}_{1^\prime}}$}
\Vertex(50,43.3){2}
\put(58,48.3){\footnotesize${ {}_{2^\prime}}$}
\Vertex(50,0){2}
\put(58,5){\footnotesize${ {}_{3^\prime}}$}
\Vertex(25,0){2}
\put(33,5){\footnotesize${ {}_{1^{\prime\prime}}}$}
\Vertex(62.5,21.65){2}
\put(70.5,26.65){\footnotesize${ {}_{2^{\prime\prime}}}$}
\Vertex(25,43.3){2}
\put(33,48.3){\footnotesize${ {}_{3^{\prime\prime}}}$}
\end{picture}
\caption{{\footnotesize \em The fixed-point structure in the $\Z_3 \times \Z_3^\prime$ model.
We label the 9 $\alpha$ fixed planes with $P_{ab\bullet}$, the
27 $\alpha\beta$ fixed points with $P_{a^\prime bc}$, the 27
$\alpha\beta^2$ fixed points with $P_{a^{\prime\prime} bc}$, and
the 3 $\beta$ fixed planes with $P_{\bullet\bullet c}$.}}
\vskip -10pt
\end{center}
\end{figure}

\subsection{Closed string spectrum}

It is convenient to consider first the massless closed string spectrum in the parent
Type IIB orbifold, before the $\Omega$ projection. In this case, we get an untwisted sector, and both
SUSY-preserving and SUSY-breaking twisted sectors.

The untwisted sector contains the 4D space-time part of the NSNS spectrum, i.e.
the graviton, the axion and the dilaton; furthermore, there are 10 scalars arising
from fields with internal indices in the NSNS sector, 12 scalars from the RR sector
and 2 spinors for each chirality from the NSR+RNS sectors.

The twists $\alpha$ and $\alpha^2$ act only on two of the three tori,
and their action preserves $N\!=\! 2$ SUSY in 6D. More precisely,
they preserve the supercharges $Q_2^{L,R}$ and $Q_3^{L,R}$ in the 4D
notation of \cite{root}. Each twisted sector contains a 6D $N\!=\! 2$ tensor multiplet.
The states are located at the $\Z_3$ fixed points, and are $\Z_3$-invariant,
while $\Z_3^\prime$ acts exchanging states from one fixed point to the other,
so that in the first torus the three $\Z_3$ fixed points are identified.

The twists $\alpha\beta$ and $(\alpha\beta)^2$ act instead on all the compact space,
preserving two supercharges, $Q_4^{L,R}$. The twisted spectrum contains a 4D $N\!=\! 2$
hypermultiplet. In these sectors, the elements $\alpha$ and $\alpha\beta^2$ act by
exchanging states from one fixed point to the other in the first torus, so that,
as before, there is only one physical fixed point in this torus.
The $\alpha\beta^2$ and $(\alpha\beta^2)^2$ twisted sectors can be treated similarly,
the only difference being the position of the fixed points and the unbroken supercharges
$Q_1^{L,R}$.

The twists $\beta$ and $\beta^2$ are SUSY-breaking, and the corresponding twisted sectors
yield each a real would-be tachyon of mass $m^2=-4/3+2T_2/(3\sqrt{3}\al )$ (where $T_2$
is the imaginary part of the K\"ahler structure of the SS torus)
and 16 massive RR 16 scalars.
These states are $\beta$-invariant and located
at $\beta$ fixed points, and again the remaining elements only switch fixed points.

It is now easy to understand the effect of the $\Omega$ projection.
In the untwisted sector, $\Omega$ removes the axion, half of the NSNS and RR scalars,
and half of the fermions. In the twisted sectors, $\Omega$ relates $Q^L$ to $Q^R$ and projects
away half of the supersymmetries, so that the surviving states fill supermultiplets of $N\!=\! 1$
SUSY in 4D or 6D. Furthermore, $\Omega$ relates the twist $\alpha \beta^{i}$ to
$(\alpha\beta^{i})^2$, and only half of the corresponding states survive the projection.
The spectrum is therefore reduced to 2 hypermultiplets of 6D $N\!=\! 1$ SUSY from
$\alpha$ twists, for each $\alpha$ fixed point; 1 chiral multiplet of 4D $N\!=\! 1$ SUSY from
$\alpha\beta$ twists, for each $\alpha\beta$ fixed point, and the same for $\alpha\beta^2$
twists; 1 real would-be tachyon and 16 massive scalars from $\beta$ twists.
The massless closed string spectrum is summarized in Table \ref{closedspectrum}.

\begin{table}[h]
\vbox{
$$\vbox{\offinterlineskip
\hrule height 1.1pt
\halign{&\vrule width 1.1pt#
&\strut\quad#\hfil\quad&
\vrule width 1.1pt#
&\strut\quad#\hfil\quad&
\vrule#
&\strut\quad#\hfil\quad&
\vrule width 1.1pt#\cr
height3pt
&\omit&
&\omit&
&\omit&
\cr
&\hfil ${\rm Sector}$ &
&\hfil  $\c\Z_6^\prime \times \Z_2^\prime$&
&\hfil  $\c\Z_3 \times \Z_3^\prime$&
\cr
height3pt
&\omit&
&\omit&
&\omit&
\cr
\noalign{\hrule
}
height3pt
&\omit&
&\omit&
&\omit&
\cr
& \hfil Untwisted&
&\hfil 1 graviton, 5 scalars&
&\hfil 1 graviton, 11 scalars, 1+1 spinors&
\cr
height3pt
&\omit&
&\omit&
&\omit&
\cr
\noalign{\hrule}
height3pt
&\omit&
&\omit&
&\omit&
\cr
&\hfil $\theta$ twisted&
&\hfil 6 chiral multiplets &
&\hfil 6 hypermultiplets &
\cr
height3pt
&\omit&
&\omit&
&\omit&
\cr
\noalign{\hrule}
height3pt
&\omit&
&\omit&
&\omit&
\cr
&\hfil $\theta^2$ twisted&
&\hfil 18 scalars&
&\hfil ${\bf -}$ &
\cr
height3pt
&\omit&
&\omit&
&\omit&
\cr
\noalign{\hrule}
height3pt
&\omit&
&\omit&
&\omit&
\cr
&\hfil $\theta^3$ twisted&
&\hfil 6 chiral multiplets  &
&\hfil ${\bf -}$ &
\cr
height3pt
&\omit&
&\omit&
&\omit&
\cr
\noalign{\hrule}
height3pt
&\omit&
&\omit&
&\omit&
\cr
&\hfil $\theta\beta$ twisted&
&\hfil 6 chiral multiplets &
&\hfil 9 chiral multiplets &
\cr
height3pt
&\omit&
&\omit&
&\omit&
\cr
\noalign{\hrule}
height3pt
&\omit&
&\omit&
&\omit&
\cr
&\hfil $\theta^3\beta$ twisted&
&\hfil 6 chiral multiplets &
&\hfil ${\bf -}$ &
\cr
height3pt
&\omit&
&\omit&
&\omit&
\cr
\noalign{\hrule}
height3pt
&\omit&
&\omit&
&\omit&
\cr
&\hfil $\theta\beta^2$ twisted&
&\hfil ${\bf -}$ &
&\hfil 9 chiral multiplets &
\cr
height3pt
&\omit&
&\omit&
&\omit&
\cr
}
\hrule height 1.1pt}
$$
}
\caption{{\footnotesize \em Massless closed string spectrum for $\Z_6^\prime \times \Z_2^\prime$
and $\Z_3 \times \Z_3^\prime$  models. We used $\theta$ as the generator
of $\Z_6^\prime$ ($\Z_3$) and $\beta$ as the generator of $\Z_2^\prime$ ($\Z_3^\prime$).
Hypermultiplets are multiplets of N=1 SUSY in 6D, while chiral
multiplets are multiplets of N=1 SUSY in 4D.
The SUSY generators are different in the different sectors, as explained
in the text.
The two spinors in the untwisted sector of $\Z_3 \times \Z_3^\prime$ have opposite
chirality.}}
\label{closedspectrum}
\end{table}

\subsection{Tadpole cancellation}

The computation of the partition functions on the $A$, $M$ and $K$ amplitudes and
the extraction of the tadpoles is again standard. The only novelty occurs in the
untwisted sector, with $\beta^n$ ($n=1,2$) inserted in the trace. In these
sectors, the oscillator contribution to the partition function is given by
\bea
\Theta_n(\tau)= \sum_{a,b=0}^{1/2} \eta_{ab} \,
\frac {{\theta{a \brack b}}^3(\tau)}{\eta^9(\tau)}\,
\frac{(-2\sin 2\pi n/3) \theta{a \brack b+2n/3}(\tau)}
{\theta{1/2 \brack 1/2+2n/3}(\tau)} \;,
\eea
and the corresponding partition function on each surface reads:
\bea
Z_{A}\raisebox{0pt}{${1 \brack \beta^n}$} \a=\a
\frac{v_4}{2NN^\prime}\int_0^\infty\frac{dt}{64t^3}
\sum_{m} e^{2i\pi n(\delta\cdot m_1)}\, \Lambda_1[m] \,
\Lambda_2[m] \Theta_{n}(it)\, ({\bf Tr}\,\delta_{n})^2 \;, \nn \\
Z_{M}\raisebox{0pt}{${1 \brack \beta^n}$} \a=\a
-\frac{v_4}{8NN^\prime}\int_0^\infty\frac{dt}{4t^3}
\sum_{m} e^{2i\pi n(\delta\cdot m_1)}\, \Lambda_1[m]\, \Lambda_2[m]
\Theta_n\left(it-1/2\right)\,{\bf Tr}\,\delta_{2n} \;, \nn \\
Z_{K}\raisebox{0pt}{${1 \brack \beta^n}$} \a=\a
\frac{v_4}{2NN^\prime}\int_0^\infty\frac{dt}{4t^3}
\sum_{m} e^{2i\pi n(\delta\cdot m_1)}\, \Lambda_1[{m}/{\sqrt{2}}]\,
\Lambda_2[{m}/{\sqrt{2}}] \Theta_{2n}(2it) \;,
\label{direct}
\eea
where $NN^\prime$ is the total order of the group (i.e. 9 in our case)
and $\Lambda_i[m]$ is the 2D lattice of the $i$-th torus as defined
in (\ref{LattDef}).

The tadpoles for massless closed string modes are easily computed. We skip the
explicit form of the usual 10D tadpole, arising in all orientifold models,
that fixes to 32 the number of $D9$-branes and requires $\gamma_\Omega^t = \gamma_\Omega^{}$.
All other tadpoles are associated to twisted states occurring only at fixed points or
fixed planes. We list them here using the already introduced notation.
We denote the twist matrices associated to the $\Z_3$ and $\Z_3^\prime$ actions
by $\gamma$ and $\delta$, and we assume $\gamma^3 = \eta_\gamma I$, $\delta^3 = \eta_\delta I$,
where $\eta_\gamma,\eta_\delta = \pm 1$. The tadpoles are at the 9 $\alpha$ fixed planes,
the 27 $\alpha\beta$ fixed points and the 27 $\alpha\beta^2$ fixed points; they
are given by $(1_{NSNS}-1_{RR})v_4/72\int dl$ times:
\bea
\alpha:
\a\a \frac{v_3}{3} \sum_{a,b} \Big[(8 - \eta_\gamma\,{\bf Tr}\, \gamma)^2
+ (8 - {\bf Tr} \,\gamma^2)^2 \Big] \,, \nn \\
\alpha\beta:
\a\a \frac{1}{3\sqrt{3}} \sum_{a^\prime,b,c} \Big[(4 + \eta_\gamma \eta_\delta\,{\bf Tr}\, \gamma\delta)^2
+ (4 + {\bf Tr}\, \gamma^2\delta^2)^2 \Big] \,, \nn \\
\alpha\beta^2:
\a\a \frac{1}{3\sqrt{3}} \sum_{a^{\prime\prime},b,c} \Big[(4 + \eta_\gamma\,{\bf Tr}\, \gamma\delta^2)^2
+ (4 + \eta_\delta\,{\bf Tr} \,\gamma^2\delta)^2 \Big] \;.
\label{twisttad}
\eea
We wrote explicitly the contributions from the $\theta^k$ and $\theta^{N-k}$ sectors,
arising from the same physical closed string state.

By taking the transverse channel expressions of the amplitudes (\ref{direct}) through
$S$ and $P$ modular transformations, the additional tadpoles for the non-SUSY
$\beta$ twisted sectors arising at the 3 fixed planes $P_{\bullet \bullet c}$ can be derived.
They yield the following result for the massive would-be tachyonic NSNS states:
\bea
\frac{v_1v_2}{4\sqrt{3}}\,\,q^{-\frac{1}{3}}\,\sum_{c}\sum_{m=-\infty}^\infty \Bigg\{\!\!
\a\a\hat\Lambda_1\left(2m+\frac{1}{3}\right) ({\bf Tr}\,\delta)^2+
\hat\Lambda_1\left(2m-\frac{1}{3}\right) ({\bf Tr}\,\delta^2)^2 + \label{tadz3} \\
\a\a\hat\Lambda_1\left(2m-\frac{2}{3}\right) (16 - {\bf Tr}\,\delta^2)^2+
\hat\Lambda_1\left(2m+\frac{2}{3}\right)  (16 - \eta_\delta\,{\bf Tr}\,\delta)^2
\Bigg\}\;, \nn
\eea
where we have retained the lattice sum along the SS directions. These tadpoles
are associated to massive states for sufficiently large radii along the SS torus,
and are therefore irrelevant in that case. They imply that
would-be tachyons and massive RR 7-forms are exchanged between
$D9$-branes and/or $O9$-planes. Contrarily to the $\Z_6^\prime\times \Z_2^\prime$
model, there is no choice for the twist matrix $\delta$ that makes eq.~(\ref{tadz3})
vanish.

\subsection{Open string spectrum}

In the following, we take $\eta_\gamma = \eta_\delta = 1$, because all the other choices lead to
equivalent theories. It is then easy to see that the twisted tadpoles (\ref{twisttad})
are cancelled by choosing ($\phi = \exp 2i\pi/3$):
\bea
\gamma \a = \a  {\rm diag} (I_{16}, \phi\,I_8, \phi^{-1}\,I_8) ; \nn \\
\delta \a = \a {\rm diag} (\phi\,I_4, \phi^{-1}\,I_4, I_{24}) .
\label{twistM}
\eea
Notice that the order of the entry in (\ref{twistM}) is crucial to cancel the tadpoles; the above
choice is such that $\gamma \delta^2 = \gamma_{\theta}$, where $\gamma_{\theta}$ is the twist
matrix of the 4D $N\!=\! 1$ ${\bf Z}_3$ model constructed in \cite{afiv}.

The massless open string spectrum is easily
determined. The maximal gauge group is $SO(8)\times U(8)\times U(4)$. The $U(8)\times U(4)$
factor comes from the $U(12)$ gauge factor of the 4D $N\!=\! 1$ $\Z_3$ model, which is further broken by
the $\Z_3^\prime$ projection. As in the previous model, this can be interpreted as a Wilson
line symmetry breaking. In this perspective, $\delta = I$ and $\gamma$ as above, and the
tadpoles in (\ref{twisttad}) are cancelled thanks to a (discrete) Wilson line $W$ equal to
$\delta$ along the first torus in (\ref{twistM}). Notice that all the gauginos are massive.
The spectrum of charged massless states is easily obtained and reported in Table \ref{spectrum}.

\section{Local anomaly cancellation}

Chiral string models have generically an anomalous spectrum of massless
states, but it is well known that this does not represent a problem, provided
that cubic irreducible anomalies vanish. Reducible $U(1)$ anomalies
are instead cancelled through a 4D version of the GS mechanism
\cite{GS}, and the corresponding $U(1)$ symmetries are spontaneously broken \cite{dsw}.
For Type IIB orientifold models, the absence of irreducible anomalies is
ensured by the cancellation of RR tadpoles \cite{tad/ano}; the GS mechanism
taking care of reducible anomalies is mediated by twisted RR axions \cite{iru,ssd4}.
For the models constructed in the previous sections, the same situation occurs.

\begin{table}[h]
\vbox{
$$\vbox{\offinterlineskip
\hrule height 1.1pt
\halign{&\vrule width 1.1pt#
&\strut\quad#\hfil\quad&
\vrule width 1.1pt#
&\strut\quad#\hfil\quad&
\vrule#
&\strut\quad#\hfil\quad&
\vrule width 1.1pt#\cr
height3pt
&\omit&
&\omit&
&\omit&
\cr
&\hfil Model &
&\hfil  $\c\Z_6^\prime \times \Z_2^\prime$&
&\hfil  $\c\Z_3 \times \Z_3^\prime$&
\cr
height3pt
&\omit&
&\omit&
&\omit&
\cr
\noalign{\hrule
}
height3pt
&\omit&
&\omit&
&\omit&
\cr
&\hfil $G_9:$&
&\hfil$\eqalign{U(4)^2 \times U(8)}$&
&\hfil$\eqalign{SO(8) \times U(8)\times U(4)}$&
\cr
&\hfil $G_5=G_{\5}:$&
&\hfil$\eqalign{    U(2)^2 \times U(4)}$&
&\hfil ${\bf -}$&

\cr
height3pt
&\omit&
&\omit&
&\omit&
\cr
\noalign{\hrule}
height3pt
&\omit&
&\omit&
&\omit&
\cr
&\hfil 99 scalars&
&$\eqalign{
 {\bf (4,4,1),\;(\overline 4,\overline 4,1),\;(1,1,28),}\cr
 {\bf (1,1,\overline{28}),\;(6,1,1),\;(1,4,\overline 8),}\cr
 {\bf (\overline 4,1,8),\; (1,\overline 6,1),\;(\overline 4,4,1),}\cr
 {\bf (4,1,8),\;(1,\overline 4,\overline 8)}}$&
&$\eqalign{
2{\bf (8,8,1),\;}2{\bf (1,\overline {28},1),}\cr
{\bf (8,1,4),\;(1,1,\overline 6)}
}$&
\cr
height3pt
&\omit&
&\omit&
&\omit&
\cr
\noalign{\hrule}
height3pt
&\omit&
&\omit&
&\omit&
\cr
&\hfil 99\,\, fermions&
&\hfil ${\bf -}$&
& $\eqalign{
2{\bf (8,1,4),\;} 2{\bf (1,1,\overline 6),} \cr
{\bf (1,\overline {8},4),\;}
{\bf (1,\overline 8,\overline4)\;}\cr
}$   &
\cr
height3pt
&\omit&
&\omit&
&\omit&
\cr
\noalign{\hrule}
height3pt
&\omit&
&\omit&
&\omit&
\cr
&\hfil $55$ chiral mult.&
&$\eqalign{
 {\bf (2,2,1),\;(\overline 2,\overline 2,1),\;(1,1,6),}\cr
 {\bf (1,1,\overline 6),\;(1_A,1,1),\;(1,2,\overline 4),}\cr
 {\bf (\overline 2,1,4),\; (1,\overline 1_A,1),\;(\overline 2,2,1),}\cr
 {\bf (2,1,4),\;(1,\overline 2,\overline 4)}}$&
&\hfil ${\bf -}$&
\cr
height3pt
&\omit&
&\omit&
&\omit&
\cr
\noalign{\hrule}
height3pt
&\omit&
&\omit&
&\omit&
\cr
&\hfil $95$ chiral mult.&
&$\eqalign{
 {\bf (\overline 4,1,1;\overline 2,1,1),\;(1,4,1;1,2,1),}\cr
 {\bf (4,1,1;1,1,\overline 4),\;(1,1,\overline 8;2,1,1),}\cr
 {\bf (1,\overline 4,1;1,1,4),\;(1,1,8;1, \overline 2 ,1)}}$&
&\hfil ${\bf -}$&
\cr
height3pt
&\omit&
&\omit&
&\omit&
\cr
}
\hrule height 1.1pt}
$$
}
\caption{{\footnotesize \em Massless open string spectrum for $\Z_6^\prime \times \Z_2^\prime$
and $\Z_3 \times \Z_3^\prime$  models. In the 55 sector, chiral multiplets
in the representation of $G_5$ are reported.
The matter content of the $\5\5$ sector is the same as in the $55$
sector, but in conjugate representations of $G_{\5}=G_5$.
In the 95 sector, chiral multiplets are present in representations of
$G_9\times G_5$. Again, the matter content in the $9\5$ sector
is obtained from that in the 95 sector by conjugation.}}
\label{spectrum}
\end{table}

The above considerations apply to 4D anomalies, which are determined by the massless
fields. In theories with compact extra-dimensions, however, also massive KK modes
can contribute non-vanishing anomalies, which are localized at the orbifold fixed points
and vanish globally, when integrated over the internal manifold \cite{sssz}. In fact,
global anomaly cancellation is not sufficient to guarantee the consistency of the
theory and anomalies must therefore cancel locally, i.e. point by point in the
internal space. From a 4D effective field theory point of view, this is due to the fact
that localized anomalies vanishing only globally lead to effective operators that, although
suppressed by the compactification scale, violate 4D gauge invariance. In the light of this
observation, we will study below how {\em local} anomaly cancellation is achieved in generic
orientifold models, with a detailed analysis for the particular models discussed in sections
2 and 3. Interestingly, this will also allow us to clarify a point that was left partially
unsolved in \cite{ssd4}, concerning the factorization of anomalous couplings in twisted sectors
with fixed planes; see Appendix B.

As expected, we find that all anomalies are cancelled through a GS mechanism. An important
novelty occurs however for globally vanishing anomalies, that correspond to an anomaly
polynomial\footnote{We use here and in the following the standard characterization of anomalies
in $D$ dimensions through a polynomial of curvatures with total degree $D+2$. The anomaly itself
is given by the Wess--Zumino descent of $I$ (see e.g. \cite{ghm}).}
$I$ that vanishes when integrated over the orbifold: $\int I = 0$.
In this case, the GS mechanism can be mediated
not only by RR axions (or their dual 2-forms), as for globally non-vanishing anomalies, but also
by KK modes of RR 4-forms. The occurrence of one or the other mechanism depends on the way the anomaly
is factorized in terms of forms $X_n(F,R)$ of definite even degree $n$, constructed out of the gauge
and gravitational curvature 2-forms $F$ and $R$.
If it has the form $I \sim X_2 X_4$, the GS mechanism will be mediated by twisted RR
axions, arising at the fixed points (or fixed-planes) where the anomaly is distributed. 
If it has instead the form
$I \sim X_6$, the relevant fields are twisted RR 4-forms, arising at fixed planes that contain all the
fixed points where the anomaly is distributed. Notice that localized irreducible anomalies are
always of the second type, whereas mixed $U(1)$ anomalies can be of both types. As we shall now
illustrate with simple and general examples, the fate of the symmetry suffering from a globally
vanishing anomaly is radically different in the two alternative mechanisms.

Consider first anomalies of the type $I \sim X_2 X_4$. In this case, the relevant GS mechanism can
be easily understood by distinguishing anomalous couplings localized at different points in
the internal space. The qualitative novelty can be illustrated by focusing on the case of a 
$U(1)$ gauge anomaly distributed at two distinct fixed points $z=z_{1,2}$,
corresponding to a term of the type $I = X_2 X_4 |_{z_1} - X_2 X_4 |_{z_2}$ in the anomaly
polynomial. This anomaly is cancelled through a GS mechanism mediated by two axions, $C_0^{1,2}$,
living at the two fixed points $z_{1,2}$\footnote{The same basic mechanism works for 
anomalies localized on fixed-planes, that will thus
be cancelled by RR axions propagating in 4 or 6 dimensions.}.
The action is:
\bea
S_{GS} \a=\a \int \! d^4 x\, \Big[\frac 12 |d C_0^1 + X_1|^2 + C_0^1 X_4 \Big]_{z_1}
+ \int \! d^4 x\, \Big[\frac 12 |d C_0^2 + X_1|^2 - C_0^2 X_4 \Big]_{z_2} \;,
\label{GS0form}
\eea
where the 1-form $X_1$ denotes the $U(1)$ gauge field
associated to the curvature 2-form $X_2$, such that $X_2 = d X_1$.
The modified kinetic terms in (\ref{GS0form}) require that $\delta C_0^{1,2} = - X_0(z_{1,2})$
under a gauge transformation\footnote{In our set-up the gauge fields that can 
have anomalies localized at distinct
fixed points are in general linear combinations of fields
coming from $D9$, $D5$ and $\bar D5$-branes.}
with parameter $X_0(x,z)$,
under which $\delta X_1 = d X_0$. The variation of the $X_4$ terms in (\ref{GS0form})
then provides the required inflow of anomaly that restores gauge invariance.
The form of the action (\ref{GS0form}) is fixed by the requirement of having {\em full} 
gauge invariance, and implies that the $U(1)$ field becomes massive, independently of
whether the anomaly vanishes or not globally. Indeed, one can choose a gauge in which
$C_0^1= -C_0^2$, where the kinetic terms in (\ref{GS0form}) are diagonalized and mass terms for
the 4D gauge field are generated. This fact has not been appreciated so far in the literature,
where only integrated anomalies were studied.

Consider next the case of anomalies of the type $I \sim X_6$. A globally non-vanishing anomaly
of this kind, associated to a global tadpole for a RR 4-form, would lead to an inconsistency,
because it cannot be cancelled by a standard GS mechanism; indeed, the latter should be mediated by a
RR 4-form, that in 4D is a non-propagating field,
whose dual in 4D would be a manifestly non-physical and meaningless
$(-2)$-form \cite{polcai}. 
Instead, if the anomaly is globally vanishing, and therefore associated
to a local tadpole for a RR 4-form, the situation is different. The crucial observation is that
this type of anomaly always appears in conjunction with twisted RR states living on fixed planes
rather than fixed points in the internal space. Such states propagate in 6D rather than 4D, and
this opens up new possibilities, since
a 4-form is now a physical propagating field and can mediate a GS mechanism.
Moreover a 4-form in 6D is dual to a 0-form, and not to a meaningless $(-2)$-form.
However, internal derivatives will play a role and the corresponding states will thus be massive
KK modes from the 4D point of view.
The situation is most conveniently illustrated with a simple example consisting of an
irreducible term in the anomaly polynomial of the form $I = X_6 |_{z_1} - X_6 |_{z_2}$, where the
points $z_1$ and $z_2$ differ only in the fixed-plane direction. The relevant 6D action for
the RR 4-form $C_4$ responsible for the inflow is:
\be
S_{GS} = \int\! d^6 x\, \frac 12 |d C_4 + X_5|^2
+ \int \! d^4x \,C_4\Big|_{z_1} - \int \! d^4x \, C_4 \Big|_{z_2} \;,
\label{GS4form}
\ee
where $X_5$ is the Chern--Simons 5-form associated to $X_6$, such that $X_6 = d X_5$.
The kinetic term in (\ref{GS4form}) requires that $\delta C_4 = - X_4$ under a 10D gauge
transformation, where $X_4$ is defined as usual from the gauge variation
of $X_5$: $\delta X_5 = d X_4$. The variation of the second and third terms in (\ref{GS4form}) then
provides the required inflow of anomaly\footnote{In short, localized irreducible 6-form terms in
the 4D anomaly polynomial look like reducible terms in a 6D anomaly polynomial, given by the
product of the 6-form term and a field-independent $\delta$-function 2-form.}.
Contrarily to the previous case, no $U(1)$ gauge factor is broken by (\ref{GS4form}).
Since $C_4$ enters in (\ref{GS4form}) only through massive KK states, it is interesting to
understand its effect in the 4D low-energy effective field theory. In order to do that, notice
that the 2-form $\delta(z\!-\!z_1) - \delta(z\!-\!z_2)$ can be written locally as
$d \eta(z)$ for some 1-form $\eta(z)$. Equation (\ref{GS4form}) can then be interpreted as a 6D action
with Lagrangian $L_{GS} = \frac 12 |d C_4 + X_5|^2 - \eta d C_4$. We can now integrate out the
massive modes of $C_4$ and evaluate their action on-shell. This is easily done by substituting
back into the Lagrangian the equations of motion for $C_4$, that imply $dC_4 + X_5 = {}^*\eta$
(where ${}^*$ denotes the 6D Hodge operator); it yields
$L_{GS}^{\rm eff} = -\frac 12 |\eta|^2 + \eta X_5$. Finally, we obtain the local 6D
Chern--Simons term
\be
S_{GS}^{\rm eff} = \int\! d^6 x\, \eta X_5 \;.
\label{CS}
\ee
Note that this gives, at it should, the same gauge variation as the original action,
since $\delta(\eta X_5) = \eta d X_4$, which gives
$- d\eta X_4 = -(\delta(z-z_1) - \delta(z-z_2))X_4$ after integration by parts.
Moreover, the discontinuous coefficient
$\eta$ is achieved exactly as proposed in \cite{bkp}, the only difference being that the
involved 4-form is a dynamical field in the full 10D theory, which behaves like an auxiliary
field only in the 4D effective theory. Importantly, the results of \cite{bkp} ensure that
the term (\ref{CS}) is compatible with local supersymmetry at the fixed points.

Summarizing, it is clear that there is an important qualitative difference between anomalies that
vanish globally and other that do not. From a purely 4D effective field theory point of view,
the condition $\int I = 0$ on the anomaly polynomial $I$ guarantees that the corresponding
anomaly can be cancelled through the addition of a local Chern--Simons counterterm with a
discontinuous coefficient\footnote{For example, an orbifold field theory that is globally free of
anomalies can be regulated in a gauge-invariant way by adding heavy Pauli--Villars fields with
mass terms that also have a discontinuous coefficient; the appropriate Chern--Simons term is then
automatically generated when integrating out the regulator \cite{bccrs,pr}.}. In open string models,
however, anomalies with $I \sim X_2 X_4$ always lead to a spontaneous symmetry breaking, and
only those with $I \sim X_6$ are cancelled through a local counterterm. It would be interesting
to understand whether there is some deeper physical principle determining this distinction, besides
factorization properties.

All the above considerations apply qualitatively to any orientifold model. For 6D SUSY models,
for instance, part of the GS mechanism is mediated by untwisted RR forms, and these can play the same
role as 6D twisted sectors in 4D models. In particular, we have verified local anomaly cancellation
in the SUSY 6D $\Z_2$ model of \cite{AP,GP}. In the case of maximal unbroken gauge group with all $D5$-branes
at a same fixed point, irreducible ${\rm Tr}\, F_9^4$ and ${\rm Tr}\, R^4$ terms in the
anomaly polynomial do not vanish locally and are indeed cancelled by a local GS mechanism similar to that
described in (\ref{GS4form}), but mediated by untwisted RR 6-forms propagating in 10D. Again, from
a 6D effective theory point of view, these amount to a local Chern--Simons term.

Let us now be more concrete and apply the general arguments outlined above to the
$\Z_6^\prime\times \Z_2^\prime$ and $\Z_3\times \Z_3^\prime$ models. We will begin with the
$\Z_3\times \Z_3^\prime$ model, which does not have irreducible anomalies at all, and then
analyse the more complicated $\Z_6^\prime\times \Z_2^\prime$ model, where some are present.
Fortunately, the techniques of \cite{ssd4} can be easily generalized
to study the local structure of anomalies, and some details of the analysis are reported in
Appendix B. It is convenient to define $\delta_{abc}$ as a 6D Dirac $\delta$-function in
the internal orbifold, localized at the fixed point with positions labelled $a$, $b$ and $c$
in the three $T^2$'s respectively, as reported in Figs. 1 and 3. We also define $\delta_{ab\bullet}$
as a 4D $\delta$-function in the internal space, localized at the fixed planes
with positions $a$ and $b$ in the first two $T^2$'s, and similarly for $\delta_{a\bullet c}$ and
$\delta_{\bullet bc}$. Moreover, we will denote by $F_i^\alpha$ the field strength of the $i$-th
factor of the gauge group, ordered as in Table \ref{spectrum} ($i=1,2,3$ in all cases), in the $\alpha$
(9, 5 or $\bar 5$) $D$-brane sector, and with ``tr'' the traces in fundamental representations of
the gauge groups.

\subsection{$\Z_3 \times \Z_3^\prime$ model}

The anomaly polynomial for the $\Z_3 \times \Z_3^\prime$ model is easily computed, and is encoded
in eqs.~(\ref{a1}) and (\ref{a2}). Its explicit form is given by ($\mu = 3^{-7/4}$):
\bea
I = \mu^2 \sum_{a,b,c}
\Big[\delta_{a^{\prime}bc} X_2^9 \big(X_4^9 \!+\! 4 Z_4\big)
- \delta_{a^{\prime\prime}bc} Y_2^9 \big(Y_4^9 \!+\! 4 Z_4\big)\Big] \;.
\label{anoZ3}
\eea
The quantities $X^\alpha_n$, $Y^\alpha_n$ and $Z_n$ are combinations of curvatures with
total degree $n$, and are obtained by expanding the topological charges of $D$-branes and
fixed points as defined in eqs.~(\ref{Q1})--(\ref{Q3}). Their explicit expressions are
($m_i=(1,1,0)$, $n_i=(1,-1,0)$, $s_i=(1,-1,-1)$):
\bea
\a\a X^9_2 = -\,\sqrt{3}\,m_i\,{\rm tr}F_i^\alpha \;,\;\;
Y^9_2 = - \sqrt{3}\,n_i\,{\rm tr}F_i^{\alpha} \;,\\
\a\a X^9_4 = Y^9_4 = - \frac 12\, \Big[
s_i\,{\rm tr}F_i^{\alpha\,2}\!
+ \frac 1{12}\,{\rm tr}R^2\Big] \;,\;\;
Z_4 = - \frac 1{192}\,{\rm tr}R^2 \;. \raisebox{19pt}{}
\eea
There are two anomalous combinations of $U(1)$ factors, $X_1^9$ and
$Y_1^9$, defined by $X_2^9 = d X_1^9$, $Y_2^9 = d Y_1^9$. These
have opposite anomalies at the two types of fixed points. This means that
the combination $X_1^9 - Y_1^9$ has true 4D anomalies, whereas
$X_1^9 + Y_1^9$ suffers only from a globally vanishing anomaly of the type
corresponding to (\ref{GS0form}). It can easily be verified that the integrated
anomaly coincides with the contribution of the massless chiral fermions in
the representations reported in Table \ref{spectrum}.

The leading order couplings (arising from the disk and cross-cap surfaces)
that are responsible for the cancellation of these anomalies are easily obtained
from the anomalous couplings for $D9$-branes and fixed points, eqs.~(\ref{b1}) and (\ref{b2}).
One finds:
\bea
{\cal L} \a=\a \mu \sum_{a,b,c} \delta_{a^{\prime}bc}
\Big[-d\chi_{a^{\prime}bc}\cdot X_1^9 - \chi_{a^{\prime}bc} \big(X_4^9 + 4\,Z_4\big)\Big] \nn \\
\a\;\a + \mu \sum_{a,b,c} \delta_{a^{\prime\prime}bc}
\Big[-d\chi_{a^{\prime\prime}bc}\cdot Y_1^9 + \chi_{a^{\prime\prime}bc} \big(Y_4^9 + 4\,Z_4\big)\Big] \;,
\label{gsZ3}
\eea
The first term in each row corresponds to the cross-term in a mixed kinetic term
of the form (\ref{GS0form}) for the two axions.

\subsection{$\Z_6^\prime \times \Z_2^\prime$ model}

The complete anomaly polynomial of the $\Z_6^\prime \times \Z_2^\prime$ model is encoded
in a compact form in eqs.~(\ref{c1}) and (\ref{c2}), which also distinguish between the various
$D$-brane sectors. It can be written explicitly as ($\mu = (12)^{-3/4}$):
\bea
I \a=\a \mu^2 \sum_{c=1}^3\bigg\{
2\sum_{b=1}^2 \delta_{1bc}
\Big[X_2^5 \big(\!-\!2 X_4^5 \!+\! X_4^9\big) + X_2^9 X_4^5
+ Y_2^5\,\big(Y_4^9 \!+\! 8 Z_4\big) + Y_2^9 Y_4^5 - 4 Y_6^9 \Big] \nn \\
\a\;\a \hspace{37pt} - 2 \sum_{b=1}^2 \delta_{1b^{\prime}c}
\Big[X_2^{\bar 5} \big(\!-\!2 X_4^{\bar 5} \!+\! X_4^9\big) + X_2^9 X_4^{\bar 5}
+ Y_2^{\bar 5} \big(Y_4^9 \!+\! 8 Z_4\big) + Y_2^9 Y_4^{\bar 5} - 4 Y_6^9 \Big] \nn \\
\a\;\a \hspace{37pt} + \sum_{b=1}^4
\Big(\delta_{1bc} - \delta_{1b^{\prime}c}\Big)
\Big[\!-\!X_2^9 X_4^9 + 4 Y_2^9 Z_4 + 4 Y_6^9\Big] \bigg\} \;,
\label{anoZ6}
\eea
in terms of the components of the charges (\ref{Q1})--(\ref{Q3}), which read in this case
($p_i=(1,1,2)$, $q_i=(1,1,-2)$, $r_i=(1,-1,0)$):
\bea
\a\a X_2^\alpha = -\,p_i\,{\rm tr}F_i^\alpha \;,\;\;
Y_2^\alpha = - \sqrt{3}\,r_i\,{\rm tr}F_i^{\alpha} \;,\\
\a\a X_4^\alpha = - \frac {\sqrt{3}}2\,r_i\,{\rm tr}F_i^{\alpha\,2} \;,\;\;
Y_4^\alpha = - \frac 12\,\Big[q_i\,{\rm tr}F_i^{\alpha\,2}\!
+ \frac {\alpha\!-\!1}{48}\,{\rm tr}R^2\Big] \;,\;\;
Z_4 = - \frac 1{192}\,{\rm tr}R^2 \;,\;\;\;\; \raisebox{21pt}{}\\
\a\a Y_6^\alpha = - \frac {r_i}{2\sqrt{3}}\, \Big[
{\rm tr}F_i^{\alpha\,3}\! + \frac 1{36} {\rm tr}F_i\,{\rm tr}R^2 \Big] \;.
\eea
When integrated over the internal space, eq.~(\ref{anoZ6}) is in
agreement\footnote{In particular, it reproduces the results of
\cite{antoniadis} for gauge anomalies, apart from irrelevant chirality
conventions.} with the contribution of the massless chiral fermions in the
representations listed in Table \ref{spectrum}.
There are however additional anomalies (of all types, including irreducible terms)
that do not involve the gauge fields associated to the $D5$ or $\bar D 5$ branes,
which are distributed with opposite signs at different fixed points and are therefore
not detectable in the 4D effective theory. These anomalies are generated
by KK modes of charged fields in the 99 sector. In total, there are 4 truly
anomalous $U(1)$'s, two $U(1)$'s that have only localized anomalies, and localized
irreducible anomalies.

The anomalous couplings for the two kinds of twisted axions $\chi$ and $\tilde \chi$
and the twisted 4-form $c$ are easily deduced from eqs.~(\ref{d1})--(\ref{d3}).
Defining for convenience the combination of Kronecker $\delta$-functions
$\delta_b = \delta_{b,1} + \delta_{b,2}$, we find:
\bea
{\cal L} \a=\a \mu \sum_{c=1}^3 \sum_{b=1}^4 \delta_{1bc}
\Big[-d\chi_{1bc} \cdot (X_1^9 - 2\,\delta_b\, X_1^5)
+ \chi_{1bc} (X_4^9 - 2\,\delta_b\, X_4^5) \nn \\
\a\;\a \hspace{65pt} -\, d \tilde \chi_{1 \bullet c}\cdot (2\,\delta_b\,Y_1^5)
+ \chi_{1 \bullet c} (4\,Z_4 + 2\,\delta_b\,Y_4^5) + {c}_{1 \bullet c} (4 - 8\,\delta_b)\Big] \nn \\
\a\;\a + \mu \sum_{c=1}^3 \sum_{b=1}^4 \delta_{1b^{\prime}c}
\Big[-d\chi_{1b^{\prime}c}(X_1^9 - 2\,\delta_{b^\prime}\, X_1^{\bar 5})
- \chi_{1b^{\prime}c} (X_4^9 - 2\,\delta_{b^\prime}\, X_4^{\bar 5}) \nn \\
\a\;\a \hspace{75pt} + d \tilde \chi_{1 \bullet c}\cdot(2\,\delta_{b^\prime}\, Y_1^{\bar 5})
- \chi_{1 \bullet c} (4\,Z_4 + 2\,\delta_{b^\prime}\, Y_4^{\bar 5})
- {c}_{1 \bullet c} (4 - 8\,\delta_{b^\prime})\Big] \nn \\
\a\;\a + \mu \sum_{a=1}^3 \sum_{c=1}^3 \delta_{a \bullet c}
\Big[d\chi_{a \bullet c} \cdot Y_1^9
- \tilde \chi_{a \bullet c} (Y_4^9 + 8\,Z_4) - d {c}_{a \bullet c} \cdot Y_5^9\Big] \;.
\eea
The terms relevant to the cancellation of localized irreducible anomalies are the last terms
of each square bracket. The other terms, instead, are relevant to the cancellation of reducible
$U(1)$ anomalies.

\section{Discussion and conclusions}

In this paper, two chiral 4D open string models with SS SUSY breaking have been
constructed as geometric freely acting orbifolds. In this setting, we derived
the known $\Z_6^\prime\times \Z_2^\prime$ model and constructed a new and very simple
$\Z_3\times \Z_3^\prime$ model. Both are classically stable, since all massless
NSNS and RR tadpoles vanish. The compactification backgrounds are non-SUSY deformations
of usual Calabi--Yau orbifolds. In the $\Z_3\times \Z_3^\prime$ model, the deformation
is induced by the $\Z_3^\prime$ element, which is a diagonal translation in a torus
together with a non-SUSY rotation along another torus. This deformation is very
similar to the one that gives rise to Melvin space-time backgrounds,
where a generic rotation along a non-compact plane is performed together with
a $2\pi R$ translation along a circle \cite{dgkt}\footnote{See \cite{Dmelvin} for a discussion
of $D$-branes on Melvin backgrounds.}. It would be interesting to analyse this analogy better. 
The quantum stability of both orientifolds
remains an open question that deserves further analysis.

A detailed study of local anomaly cancellation in the two models has been performed.
All pure gauge and mixed gauge-gravitational anomalies cancel, thanks to a generalized
GS mechanism that involves also twisted RR 4-forms, necessary to cancel localized
irreducible 6-form terms in the anomaly polynomial, which vanish only globally.
The 4D remnant of this mechanism is a local Chern--Simons term. The local (and global)
cancellation of reducible anomalies is instead ensured by twisted RR axions. In the
latter case, even $U(1)$ gauge fields affected by anomalies that vanish only globally in 4D
are spontaneously broken by the GS mechanism. 

Although we have not performed any detailed analysis of local anomaly cancellation
in closed string models, we believe that irreducible anomalies should be absent in that
case, whereas reducible ones might present some new feature.

\vskip 25pt
\noindent
{\Large \bf Acknowledgements}
\vskip 10pt

\noindent
We would like to thank L. Alvarez-Gaum\'e, I.~Antoniadis, C.~Angelantonj, P.~Creminelli, R.~Contino,
R.~Rabad\`an, R.~Rattazzi, L. Silvestrini, A.~Uranga and F. Zwirner for useful discussions.
This work was partially
supported by the EC through the RTN network ``The quantum structure of space-time and the
geometric nature of fundamental interactions'', contract HPRN-CT-2000-00131.

\appendix

\section{Lattice sums}
We denote the 2D lattice sum over the $i$-th torus by:
\bea
\label{lattice} \Lambda_i(\tau) = \sum_{n,m}\Lambda_i[m,n](\tau)
= \sum_{m, n} \,
q^{\frac 12 |P_L^{(i)}|^2}
\,\bar q^{\frac 12 |P_R^{(i)}|^2},
\eea
where $q=\exp[2i\pi\tau]$ and the lattice momenta are given by
\bea
P_L^{(i)} \a=\a \frac 1{\sqrt{2\,{\rm Im}\,T_i\,{\rm Im}\, U_i}}
\Big[\!- m_1\, U_i + m_2 +  T_i\,\Big(n_1 + n_2U_i\Big)
\Big] \;, \nn \\
P_R^{(i)} \a=\a \frac 1{\sqrt{2\,{\rm Im}\,T_i\,{\rm Im}\, U_i}}
\Big[\!- m_1\, U_i + m_2 + \bar T_i\, \Big(n_1 + n_2U_i\Big)
\Big] \;,
\eea
in terms of the standard dimensionless moduli $T_i$ and $U_i$,
parametrizing respectively the K\"ahler and complex structure
of the torus. We also define:
\be
\Lambda_i[m]\equiv\Lambda_i[m,0](it)\;,\;\;
\Lambda_i[w]\equiv\Lambda_i[0,w](it) \;,
\label{LattDef}
\ee
and denote respectively by $\hat\Lambda_i[m]$ and $\hat\Lambda_i[w]$
the corresponding Poisson resummed lattice sums, where the dependence on the
transformed modular parameter $l$ is understood.

In the following, we show in some detail how a translation affects the
toroidal lattice sums defined over the annulus, M\"obius strip and Klein bottle
world-sheet surfaces. In the $\Z_3\times \Z_3^\prime$ model, the translation acts
diagonally on the torus, as in \cite{root}, whereas in the
$\Z_6^\prime\times \Z_2^\prime$ model it actually acts non-trivially only
along a circle. The torus case has already been analysed (see for instance
\cite{kk,root}).

\paragraph{Annulus}

It is convenient to define $\Lambda[N,D\,|\,g]$
as the annulus lattice sum for Neumann (N) and Dirichlet (D) boundary conditions
(b.c.) with the insertion of the operator $g$. The only non-trivial case to be
considered is when $g=I,\delta$. The relevant Poisson resummed lattice sums are found
to be (omitting the index $i$ in $\hat \Lambda$):
\bea
&&\Lambda[N\,|\,I\,]=\sum_m \hat\Lambda[m]\, W_m^{(i)}(W_m^{(j)})^{-1}\label{ann3} \;,\\
&&\Lambda[D\,|\,I\,]=\sum_w \hat\Lambda[w]\, W_w^{(i)}\, (W_w^{(j)})^{-1} \label{ann1} \;,\\
&&\Lambda[N\,|\,\delta \,]=\sum_m \hat\Lambda[m+\delta]\,W_m^{(i)}(W_m^{(j)})^{-1} \label{ann4} \;,\\
&&\Lambda[D\,|\,\delta\,]=0\label{ann2} \;, \raisebox{11pt}{}
\eea
where $W_w^{(i)}$ encodes the position $X_i$ of the $i$-th brane along the corresponding
torus and $W_m^{(i)}$ is a generic Wilson line along the torus, parametrized by the
$\theta_i$ phase factors:
\bea
W_w^{(i)}=\exp[iw\cdot X_i/R] \;,\;\; W_m^{(i)}=\exp[im\cdot\theta_i] \;.
\eea
The sum (\ref{ann2}) vanishes because a translation has no fixed points
and hence the operator $\delta$ is not diagonal on the states.
The action of the translation in (\ref{ann4}) produces a phase in the KK modes
that, in the Poisson resummed lattice sums, gives a shift on $m$.
Notice that $D$-branes couple to all KK and winding modes.

\paragraph{M\"obius strip}

In this case, the N b.c. give lattice sums similar to
those in the annulus, since $\Omega$ does not act on KK modes.
For D b.c., the non-trivial cases are obtained when $g=R$ and $g=R\delta$, where
$R$ and $\delta$ are respectively a rotation and a translation of order 2
on the torus (actually only on a circle). Indicating with $\Lambda[N,D\,|\,\Omega g]$
the M\"obius strip lattice sum contribution, we therefore get:
\bea
&&\Lambda[N\,|\,\Omega I \,]=\sum_m \hat\Lambda[2m] \, W_{2m}^{(i)}\label{MB3}\\
&&\Lambda[N\,|\,\Omega \delta \,]=\sum_m \hat\Lambda[2m+2\delta] \, W_{2m}^{(i)}
\label{MB4} \;,\\
&&\Lambda[D\,|\,\Omega R \,]=\sum_w \hat\Lambda[2w]\, W_{2w}^{(i)} \label{MB1} \;,\\
&&\Lambda[D\,|\,\Omega R \delta \,]=\sum_w e^{2i\pi \delta \cdot w}
\hat\Lambda[2w]\, W_{2w}^{(i)} \label{MB2} \;.
\eea
The fact that only even KK and winding mode appear in the above equations
implies that $O$-planes couple only to even KK momenta and winding modes.
Notice, moreover, that eq.~(\ref{MB1}) represents the interaction of a $D5$- or
$\bar D 5$-brane with $O5$-planes in the $R$ fixed points, i.e.
$y=0$ and $y=\pi R$ along the SS direction, whereas eq.~(\ref{MB2})
represents the interaction of a
$D5$- or $\bar D 5$-brane with the $O5$-planes (actually $\bar O 5$-planes due
to the $(-)^F$ action that comes together with $\delta$) located at the $R\delta$
fixed points, i.e. $y=\pi R/2$ and $y=3 \pi R/2$ along the SS direction.
Similarly, eqs.~(\ref{MB3}) and (\ref{MB4}) represent respectively the $D9$
(or $\bar D 9$) interactions with $O9$ and $\bar O 9$-planes.

\paragraph{Klein bottle}

Define $\Lambda_i[ h \,|\, \Omega\,g]$ as the Klein bottle lattice sum
in the $h$ twisted sector with the insertion of the operator $g$
in the trace. Since lattice sums can only appear for the usual untwisted sector
or for sectors twisted by a translation of order 2, $h=I,\delta$, where
$\delta$ is the translation. On the other hand, non-trivial lattice
contributions are obtained when $g$ is a generic translation, as well
as a $\Z_2$ reflection $R$ (aside the identity).
As in the analogue annulus case, the insertion of a translation gives
rise to KK-dependent phases $\exp{(2i\pi \delta\cdot m)}$, whereas
the $\delta$ twisted sector presents half-integer winding modes
for $\Lambda_i$. Therefore, the relevant Poisson resummed lattice sums are given by:
\bea
&&\hat\Lambda[I\,|\, \Omega]=
\sum_m\hat\Lambda[2m] \;,\nn\\
&&\hat\Lambda[I \,|\, \Omega\,\delta]=
\sum_m\hat\Lambda\left[2m+2 \delta \right] \;,\nn\\
&&\hat\Lambda[ I \,|\, \Omega\, R]=
\hat\Lambda[I \,|\, \Omega\, R\,\delta]=
\sum_w\hat\Lambda[2w] \;,\nn\\
&&\hat\Lambda[ \delta \,|\, \Omega]=
\hat\Lambda[\delta \,|\, \Omega\,\delta]=0 \;,\nn\\
&&\hat\Lambda[\delta\,|\,\Omega\,R]=
\hat\Lambda[\delta\,|\,\Omega\,R\,\delta]=
\sum_w e^{2i\pi \delta \cdot w} \hat\Lambda[2w] \;. \raisebox{20pt}{}
\label{Klatt}
\eea
Notice that (\ref{Klatt}) confirms that $O$-planes couple only to even
$KK$ momenta or even winding modes, differently from $D$-branes.

\section{Anomalous couplings }

In this appendix, we discuss the computation of anomalies for the $\Z_6^\prime \times \Z_2^\prime$
and $\Z_3 \times \Z_3^\prime$ models, and the deduction of anomalous couplings by factorization.
We proceed along the lines of \cite{ssd4}. We use a compact differential form notation where
$C^\pm_{abc}$ denotes the formal sum/difference of a RR axion (0-form) $\chi_{abc}$ and its
4D dual 2-form $b_{abc}$ arising at a generic fixed point $P_{abc}$:
$C^\pm_{abc} = \chi_{abc} \pm b_{abc}$. The inflows mediated by
these fields can then be schematically written as $\langle C^\pm_{abc} C^\pm_{abc} \rangle = \pm 1$.
A similar notation is adopted also for twisted states associated to a fixed plane, say $P_{ab\bullet}$,
which consist of an axion $\chi_{ab\bullet}$ and its 6D dual 4-form $c_{ab\bullet}$, and a self-dual
2-form $b_{ab\bullet}$; we define in this case $D_{ab\bullet} = \chi_{ab\bullet} + b_{ab\bullet} 
+ c_{ab\bullet}$. Since these fields live in 6D, $D5$-branes or fixed-points and $D9$-branes or fixed-planes
couple to different 4D components $D^{9,5}_{ab\bullet}$. In particular, the 6D 2- and 4-form fields 
$b_{ab\bullet}$ and $c_{ab\bullet}$ give rise in 4D to 2- and 4-forms $b_{ab\bullet}$ and $c_{ab\bullet}$ 
when no index is in the fixed-plane direction,
but also to 0- and 2-forms $\tilde \chi_{ab\bullet}$ and 
$\tilde b_{ab\bullet}$ when 2 indices are in the fixed-plane direction.
In this notation, $D^9_{ab\bullet} = \chi_{ab\bullet} + \tilde \chi_{ab\bullet} + \tilde b_{ab\bullet}$
and $D^5_{ab\bullet} = \chi_{ab\bullet} + b_{ab\bullet} + c_{ab\bullet}$. Since $\chi_{ab\bullet}$ and 
$\tilde b_{ab\bullet}$, as well as $\tilde \chi_{ab\bullet}$ and $b_{ab\bullet}$, are dual from the 4D 
point of view, whereas $\chi_{ab\bullet}$ and $c_{ab\bullet}$ are dual from the 6D point of view, the 
only non-vanishing inflows mediated by these fields can be formally summarized in 
$\langle D^9_{ab\bullet} D^5_{ab\bullet} \rangle = 1$. 
This setting allows us to understand the form of the anomalous couplings in sectors with fixed planes, 
including those left unexplained in \cite{ssd4}.

As for standard $D$-branes \cite{ghm} and $O$-planes \cite{anomal}, it is very useful to define the
following field-dependent topological charges for $D$-branes and fixed points:
\bea
\a\a X^\alpha(F_\alpha,R) = {\rm Tr}\,[\Gamma_X^\alpha\,e^{i F_\alpha}]\,\sqrt{A(R)}
\label{Q1} \;,\\
\a\a Y^\alpha(F_\alpha,R) = {\rm Tr}\,[\Gamma_Y^\alpha\,e^{i F_\alpha}]\,\sqrt{A(R)}
\label{Q2} \;,\\
\a\a Z(R) =  \sqrt{L(R/4)} \label{Q3} \;.
\eea
The labels $X$ and $Y$ distinguish between the two different sectors contributing to the
anomaly in each of the models under analysis. These charges must be intended as sums of components
with growing degree $n$, which we shall denote by $X^\alpha_n$, $Y^\alpha_n$ and $Z_n$.

\paragraph{$\Z_3\times \Z_3^\prime$ model}

In the $\Z_3\times \Z_3^\prime$ model, $X$ refers to the $\theta\beta$ twisted sector,
whereas $Y$ refers to the $\theta\beta^2$ twisted sector, so that $\Gamma_X = \gamma\,\delta$
and $\Gamma_Y = \gamma\,\delta^2$. The anomaly polynomial is easily computed and is given by
$I = I_A^{99} + I_M^9$, where
\bea
I_A^{99} \a=\a \frac {\mu^2}{2} \sum_{a,b,c}
\Big(\delta_{a^{\prime}bc}\,X^9\,X^9
- \delta_{a^{\prime\prime}bc}\,Y^9\,Y^9\Big) \label{a1} \;,\\
I_M^{9} \a=\a 2\,\mu^2 \sum_{a,b,c}
\Big(\delta_{a^{\prime}bc}\,X^9\,Z
-\delta_{a^{\prime\prime}bc}\,Y^9\,Z\Big) \label{a2} \;,
\eea
are the contributions from the annulus and M\"obius strip surfaces respectively, and $\mu = 3^{-7/4}$.
The anomaly polynomial can be easily factorized, and yields the following anomalous couplings:
\bea
S_{D9} \a=\a \mu \int \sum_{a,b,c} \Big(
\delta_{a^{\prime}bc}\,C^-_{a^{\prime}bc}\,X^9
+ \delta_{a^{\prime\prime}bc}\, C^+_{a^{\prime\prime}bc}\,Y^9\Big) \label{b1} \;,\\
S_{F} \a=\a 4\,\mu \int \sum_{a,b,c}
\Big(\delta_{a^{\prime}bc}\, C^-_{a^{\prime}bc}\,Z
+ \delta_{a^{\prime\prime}bc}\, C^+_{a^{\prime\prime}bc}\,Z\Big) \label{b2} \;.
\eea

\paragraph{$\Z_6^\prime \times \Z_2^\prime$ model}

In the $\Z_6^\prime \times \Z_2^\prime$ model, $X$ refers to the $\theta$ and $\theta\beta$ sectors,
whereas $Y$ refers to the $\theta^2$ and $\theta^2\beta$ sectors; $\Gamma_X$ is defined as $\gamma_9$
in the $9$ sector and $\gamma_{16}$ in the $5$ and $\bar 5$ sectors, and $\Gamma_Y$ as
$\gamma_9^2$ in the $9$ sector and $\gamma_{16}^2$ in the $5$ and $\bar 5$ sectors.
The anomaly is given by $I=\sum_{\alpha\beta} I_A^{\alpha\beta} + \sum_{\alpha} I_M^\alpha$
in terms of the contributions from each sector of the annulus and M\"obius strip, which are given
by
\bea
I_A^{\alpha\beta} \a=\a - \frac {\mu^2}2 \sum_{b=1}^{n^{\alpha\beta}} \sum_{c=1}^3 \rho^{\alpha\beta}
\Big[\delta_{1bc}\,\Big(X^\alpha\,X^\beta + \epsilon^{\alpha\beta}\, Y^\alpha\,Y^\beta\Big)
-\delta_{1b^{\prime}c}\,\Big(X^{\bar \alpha}\,X^{\bar \beta} 
+ \epsilon^{\bar\alpha\bar\beta}\, Y^{\bar\alpha}\,Y^{\bar\beta}\Big)\Big]
\label{c1} \;,\;\;\;\;\; \\
I_M^{\alpha} \a=\a 4\,\mu^2 \sum_{b=1}^{n^{\alpha\alpha}} \sum_{c=1}^3 \rho^{\alpha\alpha}
\Big[\delta_{1bc}\,Y^\alpha\,Z -\delta_{1b^{\prime}c}\,Y^{\bar \alpha}\,Z\Big]
\label{c2} \;,
\eea
where $\mu = 12^{-3/4}$; for $\alpha\beta=99,55,95,59$, the coefficient
$\rho^{\alpha\beta}$ is equal to $1,4,2,2$, $\epsilon^{\alpha\beta}$ is $0,0,1,1$,
and $n^{\alpha\beta}$ is $4,2,2,2$.
Written in this form, the anomaly can be easily factorized, and we
find the following anomalous couplings:
\bea
S_{D9} \a=\a \mu \sum_{c=1}^3 \int \Big[
\sum_{b=1}^4 \Big(\delta_{1bc}\,C^+_{1bc} \,X^9 + \delta_{1b^{\prime}c}\, C^-_{1b^{\prime}c}\,X^9\Big)
- \sum_{a=1}^3 \Big(\delta_{a \bullet c}\, D_{a \bullet c}\,Y^9\Big) \Big]
\label{d1} \;, \\
S_{D5} \a=\a -2\,\mu \sum_{c=1}^3 \int \sum_{b=1}^2
\Big[\delta_{1bc}\,\Big(C^+_{1bc}\,X^5 - {D}_{1 \bullet c}\,Y^5 \Big)
+ \delta_{1b^{\prime}c}\,\Big(C^-_{1b^{\prime}c}\, X^{\bar 5}
+ {D}_{1 \bullet c}\,Y^{\bar 5}\Big) \Big] \label{d2} \;,\;\;\;\; \\
S_{F} \a=\a 4\,\mu \sum_{c=1}^3 \int \Big[
\sum_{b=1}^4 \Big(\delta_{1bc}\, {D}_{1 \bullet c}\,Z
- \delta_{1b^{\prime}c}\, {D}_{1 \bullet c}\,Z\Big)
-2 \sum_{a=1}^3 \Big(\delta_{a \bullet c}\,D_{a \bullet c}\,Z\Big)\Big] \label{d3} \;.
\eea

\end{document}